\documentclass[iop]{emulateapj}
\usepackage{amssymb,amsmath,mathtools}
\usepackage{graphicx}
\usepackage{subfigure}
\usepackage{natbib}
\usepackage[usenames,dvipsnames]{xcolor}
\usepackage{enumitem}

\newcommand \beq {\begin{equation}}
\newcommand \eeq {\end{equation}}
\newcommand \lp {\left(}
\newcommand \rp {\right)}

\newcommand \ls {\left[}
\newcommand \rs {\right]}

\newcommand \msun {\, M_{\odot}}
\newcommand \msunyr {\, \msun\, {\rm yr}^{-1}}
\newcommand \G {\, {\rm G}}
\newcommand \Gcmcube {\, \rm{G\, cm^3}}
\newcommand \gcmsq {\, \rm{g\, cm^2\, s^{-2}}}
\newcommand \ergsec {\, \rm{erg \, s^{-1}}}
\newcommand \ms {\, {\rm ms}}
\newcommand \km {\, {\rm km}}
\newcommand \Hz {\, {\rm Hz}}
\newcommand \Hzs {\, {\rm Hz\, s^{-1}}}

\newcommand \mdot {\dot{M}}
\newcommand \mdedd {\mdot_{\rm Edd, \odot}}
\newcommand \rmag {r_{\rm m}}
\newcommand \rmagmax {r_{\rm m, max}}
\newcommand \rstar {r_*}
\newcommand \rco {r_{\rm co}}
\newcommand \rlc {r_{\rm LC}}
\newcommand \ra {r_{\rm A}}

\newcommand \Nacc {N_{\rm acc}}
\newcommand \Nacctt {N_{\rm acc, 32}}
\newcommand \Nfield {N_{\rm field}}
\newcommand \Nup {N_{\rm up}}
\newcommand \Ndown {N_{\rm down}}
\newcommand \Ndowntt {N_{\rm down, 32}}

\newcommand \Ntot {N_{\rm tot}}
\newcommand \Ndownopen {N_{\rm down,open}}

\newcommand \Lj {L_{\rm j}}
\newcommand \ff {\psi}
\newcommand \ffopen {\ff_{\rm open}}
\newcommand \nueq {\nu_{\rm eqlm}}
\newcommand \No {N_0}
\newcommand \Lo {L_0}

\newcommand \ffopeno {\ff_{\rm open, 0}}
\newcommand \peq {P_{\rm eqlm}}
\newcommand \peqstd {P_{\rm eqlm, std}}
\newcommand \nudot {\dot{\nu}}
\newcommand \wfast {\omega_{\rm fast}}
\newcommand \wfasteq {\omega_{\rm fast, eqlm}}

\newcommand \numin {\nu_{\rm min}}
\newcommand \numax {\nu_{\rm max}}

\newcommand \Lx {L_{\rm X}}
\newcommand \Lr {L_{\rm R}}

\renewcommand{\vec}[1]{\boldsymbol{#1}}

\begin{document}

\title{Torque Enhancement, Spin Equilibrium, and Jet Power from \\Disk-Induced Opening of Pulsar Magnetic Fields}

\author{Kyle Parfrey, Anatoly Spitkovsky}
\affil{Department of Astrophysical Sciences, Princeton University, Peyton Hall, Princeton, NJ 08544, USA}
\email{parfrey@astro.princeton.edu} 
\and 
\author{Andrei M. Beloborodov}
\affil{Department of Physics, Columbia University, 538 West 120th Street, New York, NY 10027, USA}

\begin{abstract}
The interaction of a rotating star's magnetic field with a surrounding plasma disk lies at the heart of many questions posed by neutron stars in X-ray binaries.
We consider the opening of stellar magnetic flux due to differential rotation along field lines coupling the star and disk, using a simple model for the disk-opened flux, the torques exerted on the star by the magnetosphere, and the power extracted by the electromagnetic wind. 
We examine the conditions under which the system enters an equilibrium spin state, in which the accretion torque is instantaneously balanced by the pulsar wind torque alone.
For magnetic moments, spin frequencies, and accretion rates relevant to accreting millisecond pulsars, the spin-down torque from this enhanced pulsar wind can be substantially larger than that predicted by existing models of the disk-magnetosphere interaction, and is in principle capable of maintaining spin equilibrium at frequencies less than 1 kHz.
We speculate that this mechanism may account for the non-detection of frequency increases during outbursts of SAX J1808.4-3658 and XTE J1814-338, and may be generally responsible for preventing spin-up to sub-millisecond periods. 
If the pulsar wind is collimated by the surrounding environment, the resulting jet can satisfy the power requirements of the highly relativistic outflows from Cir X-1 and Sco X-1. In this framework, the jet power scales relatively weakly with accretion rate, $\Lj \propto \mdot^{4/7}$, and would be suppressed at high accretion rates only if the stellar magnetic moment is sufficiently low.
\end{abstract}

\keywords{magnetic fields --- accretion, accretion disks --- pulsars: general --- stars: neutron --- stars: jets --- X-rays: binaries}

\maketitle

\section{Introduction} 
\label{intro}

The study of accreting pulsars was inaugurated by the discovery of 4.8 s X-ray pulsations from Cen X-3 \citep{Giacconi:1971aa}. These first observations immediately revealed significant pulse period changes on timescales as short as hours and a secular decrease in period, characteristics that would become central to understanding this new source class. Shortly thereafter 1.2 s pulsations were found from Her X-1 \citep{Tananbaum:1972aa}, again with a gradually decreasing period \citep{Giacconi:1973aa}. These and other similar sources were soon recognized as neutron stars accreting from binary companions \citep*{Pringle:1972aa, Davidson:1973aa, Lamb:1973aa}, and the basic theory that was erected around them was found to agree approximately with their measured spin phenomenology \citep{Rappaport:1977aa}. The pulses themselves were deduced to be caused by channeling of accreting plasma to hotspots near the star's magnetic poles, with the X-rays released in the accretion shock subject to additional beaming in the inferred strong magnetic field \citep{Gnedin:1973aa, Basko:1975aa}.

Two particularly interesting classes of accreting neutron stars are the millisecond X-ray pulsars and those which power relativistic radio jets. Millisecond spin periods have been observed for both accretion-powered sources, which emit from a magnetically confined accretion column, and nuclear-powered sources whose pulses are observed during type I X-ray bursts. Their spin frequencies are consistent with a uniform distribution up to a cut-off frequency of 730 Hz \citep{Chakrabarty:2003aa, Chakrabarty:2008aa}. This is somewhat surprising since the cut-off is well below the critical break-up (mass-shedding) frequency estimated for a range of internal equations of state, about 1230 Hz \citep*[e.g.][]{Cook:1994aa,Lattimer:2004aa}. Furthermore, for several objects the standard accretion torque theory overpredicts the increase in spin frequency that should be measured over the course of an outburst given the observed X-ray flux \citep{Patruno:2012aa}, implying either that the accretion torques are less efficient than expected or that any torque acting to spin down the star is anomalously strong.

Radio jets have been detected from several binaries hosting neutron stars, most dramatically the non-pulsing sources Cir X-1 \citep{Stewart:1993aa} and Sco X-1 \citep{Geldzahler:1981aa}. Both of these outflows are truly relativistic: energy transfer from the core to the radio lobes has been estimated to occur with Lorentz factor $\Gamma > 3$ in Sco X-1 \citep*{Fomalont:2001ab} and $\Gamma > 15$ in Cir X-1 \citep{Fender:2004aa}. Extended radio emission has also been associated with several accreting millisecond pulsars, which raises the possibility that their jets may be powered by stellar rotation, analogous to models of spin-powered black hole jets. Unlike in black hole systems, there is no evidence for suppression of the jet in the soft X-ray state for most accreting neutron stars \citep*{Migliari:2006aa}, perhaps because one or more of the components of the launching process is fundamentally different.

A variety of theoretical models have been proposed to describe the interaction between the rotating neutron star's roughly dipolar magnetic field and the highly conducting accretion disk, in tandem with developments in the closely related problem of protostellar accretion; see \cite{Uzdensky:2004ab} and \cite{Lai:2014aa} for reviews. Two principal states exist, distinguished by whether the disk's inner edge is inside or beyond the corotation radius, 
\beq
\rco = \lp\frac{GM}{\Omega^2}\rp^{1/3}\,,
\label{eq:rco}
\eeq
at which the Keplerian and stellar-rotation frequencies are equal; $M$ and $\Omega$ are the star's mass and spin angular frequency. If the inner disk is inside $\rco$ it rotates faster than the star and accretion along magnetic field lines can occur. If the disk is terminated outside corotation a centrifugal barrier prevents accretion, and the rotating field lines can eject matter from the system at the expense of the star's rotational energy \citep{Davidson:1973aa, Illarionov:1975aa, Davies:1981aa}; this is known as the propeller regime.

The stellar field may penetrate the disk \citep*{Ghosh:1977aa, Ghosh:1978aa, Ghosh:1979aa}, resulting in spin-up (spin-down) torques acting on the star from field lines entering the disk inside (outside) the corotation radius; an accreting star can either gain or lose angular momentum, depending on which of the spin-up/spin-down torque contributions is larger. In this picture the field lines slip through the disk due to the effective magnetic diffusivity provided by turbulence and remain essentially dipolar out to large radii. However, even with turbulent diffusivity a thin thermal disk remains a good conductor, leading \cite{Aly:1980aa} to construct a magnetospheric solution in which surface currents successfully exclude the stellar field from the disk \citep[see also][]{Istomin:2014aa}.

Magnetic field lines which couple the star and disk are twisted due to the mismatched angular velocities of their footpoints; if the diffusivity in the disk is too low to allow sufficiently rapid slippage, the resulting build up of toroidal field causes the field lines to balloon outwards, eventually forming a nearly radial configuration in which oppositely directed field lines are separated by a current sheet, which may be unstable to reconnection \citep*{Aly:1985aa, Aly:1990aa, Lynden-Bell:1994aa, Uzdensky:2002aa, Uzdensky:2002ac}. \cite*{Lovelace:1995aa} proposed a model in which the field lines remain open to infinity without reconnecting, allowing magnetohydrodynamic (MHD) winds to extract angular momentum both from the star and the disk. In the X-wind configuration of \cite{Shu:1994aa}, the magnetic field is pinched inward and stellar field lines only intersect the disk in a narrow annulus around the corotation radius, which approximately coincides with the disk's truncation radius. 

The predicted twisting, current sheet formation, and reconnection cycle was recovered by the first generation of time-dependent MHD simulations of accretion onto magnetic stars \citep*{Hayashi:1996aa, Goodson:1997aa, Miller:1997aa}. The reconnection events and subsequently ejected plasmoids seen in the simulations were associated with the X-ray flares and optical jets observed from classical T Tauri stars. \cite*{Kato:2004aa} demonstrated the acceleration of a transrelativistic jet by the twisted magnetic tower created by the neutron star-disk system \citep{Lynden-Bell:1996aa}. Simulations of accreting stars found solutions with steady funnel flows \citep{Romanova:2002aa, Bessolaz:2008aa}, while those of rapidly rotating stars showed the expected propeller-like behavior \citep{Romanova:2004aa, Ustyugova:2006aa}. 

The equations of non-relativistic MHD, used in the simulations described above, are satisfactory when studying protostars and even neutron stars with periods of around a second, for which the most important interactions between the accreting matter and the stellar field occur well within the light cylinder,
\beq
\rlc \equiv \frac{c}{\Omega} \approx 5\, \frac{P}{1\ms}\, \rstar,
\label{eq:rlc}
\eeq
where $P$ and $\rstar$ are the stellar spin period and radius. (In all numerical estimates we use $M = 1.4 \msun$ and $\rstar = 10 \km$.) For millisecond pulsars the light cylinder is clearly very close to the inner magnetosphere where the magnetic field configuration determines the various torques the star experiences, and so in principle a relativistic treatment is required. 

Here we present idealized calculations performed in the framework commonly used to describe the extraction of rotational energy from an isolated pulsar by an electromagnetically dominated plasma wind.
We will show that, rather than providing a small correction to existing theory, the increase in the strength of the relativistic pulsar wind due to the star-disk interaction may in some cases make this the primary mechanism by which spin-down torques are applied to the star. We will discuss applications to the accretion torques inferred during millisecond pulsar outbursts, the spin distribution of the millisecond pulsar population, and the production of relativistic jets in neutron star X-ray binaries.

We outline in Section~\ref{standard} the general principles of accretion onto stars whose internal dipole field is strong enough to terminate the Keplerian accretion flow above the stellar surface, and describe several models which have been advanced to calculate the torque exerted on the star by magnetic field lines which intersect the disk. In Section~\ref{theory} we investigate the consequences of the opening of a large fraction of these star-disk coupling field lines, and develop a highly simplified model which permits concise algebraic relationships between many of the quantities of interest. Section~\ref{eqlm} is devoted to spin equilibrium, and the conditions under which it can be achieved in our toy model.  In Section~\ref{obs} we compare the behavior which would be expected under our model to observations of the spin frequencies, frequency derivatives, and jet phenomenology of neutron star X-ray binaries. Finally we summarize and discuss our results in Section~\ref{discussion}.

\section{Standard theory of accretion onto magnetic stars}
\label{standard}

\subsection{Basic principles}

A star with a sufficiently strong magnetic field will prevent the direct accretion of matter onto its surface---at the magnetospheric radius $\rmag$ the disk is truncated and the accreting plasma is directed onto the stellar magnetic field lines, along which it funnels down to a ring around the magnetic poles, or ring segment if the dipole moment is not perpendicular to the disk plane. An estimate of the truncation point is given by the Alfv\'{e}n radius $\ra$, where the magnetic and kinetic energy densities are equal; it can be approximated by equating the magnetic pressure from the stellar dipole field to the ram pressure of gas in spherical free-fall from infinity \citep{Davidson:1973aa, Elsner:1977aa},
\beq
\ra = \lp \frac{\mu^4}{2 G M \mdot^2}\rp^{1/7},
\label{eq:ra}
\eeq
where $\mu$ is the stellar magnetic dipole moment and $\mdot$ is the mass accretion rate. It is standard practice to posit the relationship
\beq
\rmag = \xi \ra,
\label{eq:rm}
\eeq
in which additional effects, such as from the non-spherical disk geometry, are subsumed into the parameter $\xi \lesssim 1$. Simulations have generally found $\xi \sim 0.5$ \citep[e.g.][]{Long:2005aa,Bessolaz:2008aa,Zanni:2013aa}. Accretion can take place only if $\rmag < \rco$.

Matter accreting from the magnetospheric radius applies a torque to the star,
\beq
\Nacc = \mdot\, l (\rmag)  \approx \mdot \sqrt{GM\rmag} ,
\label{eq:nacc}
\eeq
where $l(r)$ is the specific angular momentum at radius $r$, given by approximately Keplerian rotation; this torque is applied by magnetic field lines twisted by the inflowing plasma. Outside the accretion funnel, the stellar magnetic field is twisted forward or backward due to dragging by the disk, and any field lines which are open (i.e., do not return to the star or intersect the disk) are swept back due to the effective inertia of the electromagnetic field \citep[e.g.][]{Bogovalov:1997aa}; together, the disk-coupling flux and open flux apply a torque
\beq
\Nfield = \Nup + \Ndown
\eeq
to the star, where $\Nup$ ($\Ndown$) represent the torque contributions which would cause the spin frequency to increase (decrease). The total torque is the sum 
\beq
\Ntot = \Nacc + \Nfield.
\eeq

\subsection{Stellar torque from the disk-magnetosphere interaction}

Several models have been proposed for estimating the torque on accreting pulsars. These are often expressed in the form
\beq
\Ntot = n(\wfast)\, \Nacc,
\eeq
where $n(\wfast)$ is a dimensionless function and
\beq
\wfast = \frac{\Omega}{\Omega_{\rm K}(\rmag)} = \lp \frac{\rmag}{\rco} \rp^{3/2}
\eeq
is known as the fastness parameter \citep{Elsner:1977aa}; models described in this way are only applicable when $\rmag<\rco$. \cite{Ghosh:1979aa} proposed the form
\beq
n(\wfast) = 1.39\, \frac{1 - \wfast\ls4.03\lp1-\wfast\rp^{0.173} - 0.878\rs}{1 - \wfast},
\label{eq:gl79}
\eeq
although it was later shown that this model is internally inconsistent \citep{Wang:1987aa}. If the magnetic stress communicated by the magnetosphere is limited by its susceptibility to field line opening and reconnection, \cite{Wang:1995aa} argued that the torque should be given by
\beq
n(\wfast) = \frac{7/6 - (4/3)\wfast + (1/9)\wfast^2}{1-\wfast}.
\label{eq:wang95}
\eeq

Accretion onto millisecond pulsars was addressed by \cite*{Rappaport:2004aa}, who argued that the disk's inner edge would adjust to be near the corotation radius when the usual estimates, such as equation~(\ref{eq:ra}), imply that $\rmag>\rco$ (i.e., $\wfast > 1$). In their model, the effective accretion torque is set by the specific angular momentum at $\rco$, and the total torque applied is given by
\beq
\Ntot = \mdot \sqrt{GM\rco} + \Nfield,
\eeq
where the contribution from field lines that couple the star and disk is
\beq
\Nfield = 
  \begin{dcases}
  \frac{\mu^2}{3 \rco^3}\ls \frac{2}{3} - \frac{2}{\wfast} + \frac{1}{\wfast^2} \rs, & \wfast < 1 \\
   -\frac{\mu^2}{9\rco^3}, & \wfast > 1.
  \end{dcases}
\label{eq:rapp04}
\eeq
Both this model and equation~(\ref{eq:wang95}) are constructed using a magnetic-field pitch distribution in the disk of $B_\phi/B_z = \lp\Omega - \Omega_{\rm K}\rp/{\rm max}\lp\Omega, \Omega_{\rm K}\rp$.

\section{Opening of pulsar magnetic flux \\by an accretion disk}
\label{theory}

\subsection{Spin-down of an isolated pulsar}

In the absence of accreting plasma, an isolated star rotating about its magnetic axis loses energy at a rate
\beq
\Lo = -\No\Omega = \mu^2 \frac{\Omega^4}{c^3} \approx  \frac{2}{3c} \Omega^2 \ffopeno^2,
\label{eq:n0}
\eeq
where $\ffopeno$ is the total magnetic flux open through the light cylinder to infinity \citep{Goldreich:1969aa, Gruzinov:2005aa, Contopoulos:2005aa} and $\No$ is the torque applied to the star by the magnetosphere. We will generally restrict our attention to axisymmetric models, which should be a good approximation for stars whose spin and magnetic axes are nearly aligned. We take torques which increase the stellar rotation rate to be positive, so spin-down torques such as $\No$ are negative. For most accreting stars the torque from this electromagnetic wind is small in comparison to that due to interaction of the stellar magnetic field with the disk, and it is usually ignored. However for neutron stars with either large magnetic moments or high spin frequencies this torque ceases to be negligible, and we will argue in particular that $\No$ sets the spin-down torque scale for accreting millisecond pulsars.

As an example, take the simulations of \cite{Ustyugova:2006aa}, who measured the spin-down due to the propeller effect on accreting plasma held outside the corotation radius. Scaling their results to a neutron star with $\mu = 10^{27} \Gcmcube$ rotating at $\Omega = 5\times10^{3}\, \rm{s^{-1}}$, they report a spin-down torque of $-3.9\times10^{33}\gcmsq$. This is actually slightly weaker than the isolated-pulsar spin-down torque for these parameters, $\No = -4.6\times 10^{33}\gcmsq$.

\subsection{Proposed torque \& jet power model}

\begin{figure}
\includegraphics[width=3.4in]{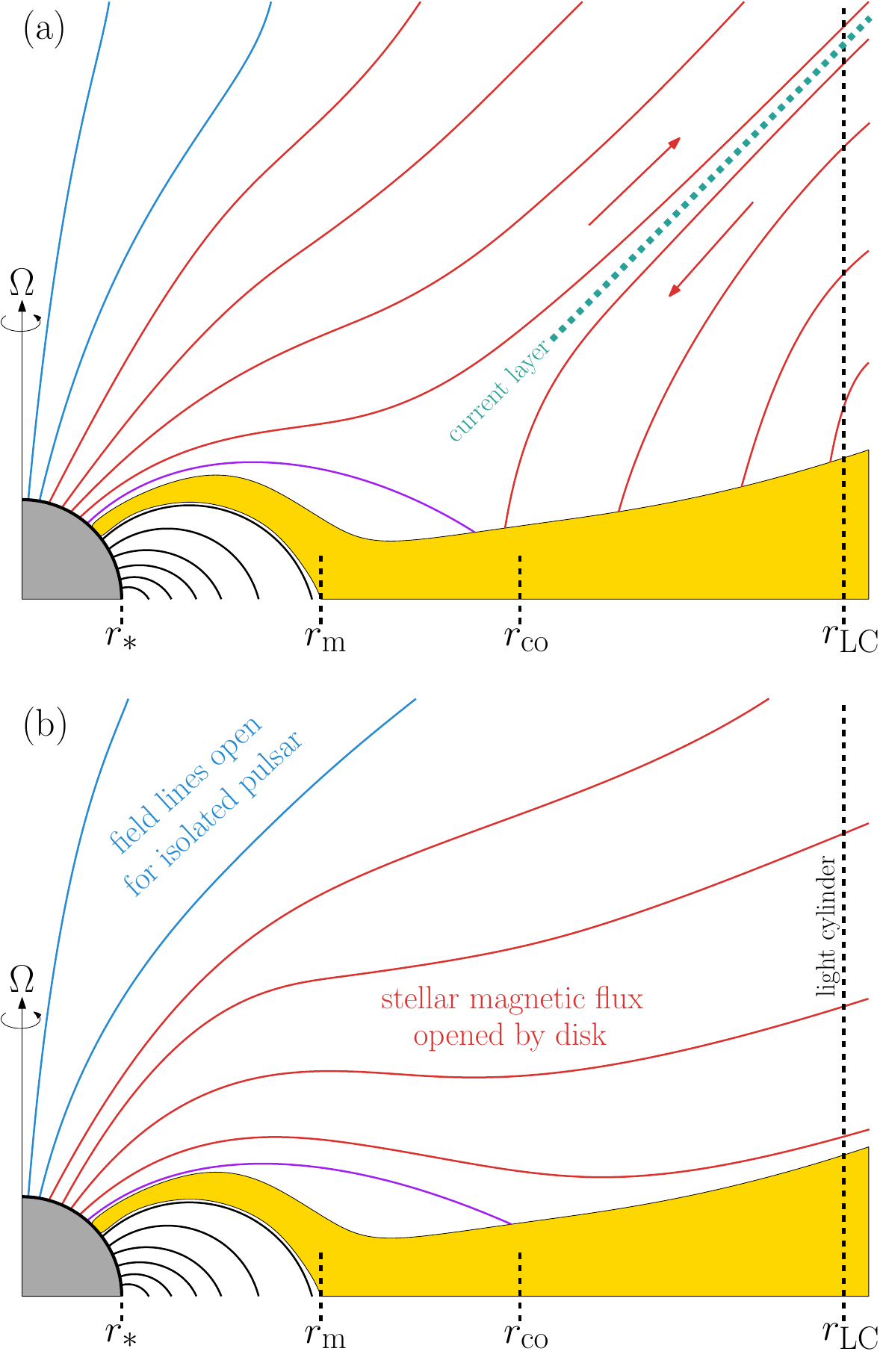}
\caption{\label{fig:cartoon} Schematic magnetic field and accretion flow geometry for the proposed model, showing the disk outside $\rmag$ and the accretion funnel inside $\rmag$ (gold region). Most stellar flux which would otherwise couple to the disk is forced to open (red lines), although a small amount of flux may be trapped near the corotation radius (purple line). Field lines which would be open in the magnetosphere of an isolated pulsar of the same spin frequency $\Omega$ are indicated in blue.  (a) Stellar field lines have opened, forming a current layer (cyan dashed line); red arrows indicate the field's poloidal direction on either side of the layer. (b) Relaxed state, in which those parts of the open stellar fleld lines which are rooted in the disk have been pushed beyond the light cylinder.}
\end{figure}

Magnetic field lines which connect the star to the disk are forced to open, unless the disk permits azimuthal field-line diffusion sufficiently rapid to prevent the accumulation of toroidal field. Here we focus on the case where azimuthal diffusion is not fast enough to inhibit opening, and construct a toy model to investigate the potential effect the additional open flux may have on various properties of the system.

In the open state, field lines of opposite direction are separated by a thin current sheet which will likely be unstable to reconnection. The average state of the system will then be determined by a competition between the rates of differential rotation (i.e., field line twisting) and reconnection. When the twisting rate is slow, reconnection of all the inflated field lines can occur before the newly reconnected flux has time to expand outward again, leading to cyclical behavior \citep*[e.g.][]{Parfrey:2012ac,Parfrey:2013aa}. In the case of accreting millisecond pulsars the differential rotation rate between the star and most points in the disk amounts to hundreds of Hz; the consequent rapid twisting can delay the onset of reconnection \citep{Parfrey:2013aa} and may lead to a long-term quasi-steady open state.

An additional consideration is that, as field lines become increasingly twisted they are driven toward a final configuration which is approximately radial in the poloidal plane. If the field lines are nearly frozen into a conducting equatorial disk they bulge outward at an angle of $\sim 60 ^\circ$ to the vertical \citep{Lynden-Bell:1994aa}---the field lines are therefore pinched radially inward where they enter the disk, and they will slip outward, relative to the disk material, on a resistive timescale \citep{van-Ballegooijen:1989aa, Bardou:1996aa, Agapitou:2000aa, Uzdensky:2002ac}. If the slippage speed is smaller than the radial accretion velocity, the segments of the opening field lines connected to the disk will move inward; if the slippage speed is greater, the field lines will diffuse outward and eventually beyond the light cylinder. 

The accretion flow and magnetic field configuration are shown in Fig.~\ref{fig:cartoon}. In Fig.~\ref{fig:cartoon}(a) a strong current layer is forming between the two sides of an outwardly inflating field line, and the opening field lines are bent forwards (or pinched) where they meet the disk; in Fig.~\ref{fig:cartoon}(b) a relaxed state has been reached, as the open field lines which are now attached solely to the disk have been pushed through the light cylinder; the state in (a) may lead to that in (b) if the effective magnetic diffusivity in the disk is high enough. We have implicitly assumed that the disk's own magnetic field is present only on small scales, where it contributes to the turbulent magnetic diffusivity by allowing the stellar field to move through the disk via reconnection; the disk field may also mediate the coupling between the stellar field and the disk.

Field lines which enter the disk very near the corotation radius are only subjected to slow twisting, and may, in isolation, be capable of slipping azimuthally fast enough to prevent the accumulation of toroidal field. However, the magnetic pressure will quickly increase on field lines entering well inside $\rco$, as these field lines experience rapid twisting; within a few radians of differential rotation the pressure will be large enough to push outward, and open, that small amount of magnetic flux coupling near the corotation radius  \citep[see e.g.][]{Parfrey:2013aa}.  For this reason, the field line crossing the equator at the corotation radius is, in general, not the same field line as would have done so in the disk's absence. In this way the disk-coupling field can be opened (and, depending on the accretion velocity, expelled through the light cylinder) all the way down to the magnetospheric radius. 

There will be some flux trapped near corotation in the steady state, as illustrated by the magenta field line in Fig.~\ref{fig:cartoon}(b); if one considers a sequence of such states with decreasing disk magnetic diffusivity, the amount of flux trapped in the disk will become infinitesimal as the disk approaches a perfect conductor. In this figure, the trapped field line entering the disk near corotation will have come from near $\rmag$ if the disk were highly conducting, and much of the newly opened (red) field lines would have initially closed inside $\rco$.

We assume that the open field lines thread a tenuous but highly conducting plasma, which for the purpose of constructing a simple model we can take to be inertialess and dissipationless (i.e., the ideal force-free approximation). Charged particles on the open field lines will move radially outward at the $\vec{E}\times\vec{B}$ drift velocity, so the plasma must be continuously supplied. Radio pulsars self-generate this plasma via magnetospheric pair production, which may also operate around accreting stars if the dense accreting plasma is not able to cross field lines onto the open flux, allowing the formation of electrostatic vacuum gaps and pair discharge. For a star of a given magnetic moment, there is a minimal spin frequency required for pair production; our sources of interest must satisfy this condition, as radio millisecond pulsars are observed with similar magnetic field strengths and rotation rates. In an alternative scenario, some of the accreting plasma may diffuse onto the open field lines, supplying the required charge carriers without the need for pair creation. 

The torque on, and power extracted from, the star are independent of the source of the wind plasma, and are insensitive to its density as long as the open flux region is magnetically dominated. The torque and power are larger if the open-zone plasma inertia is not negligible, because mass-loaded field lines are swept back or twisted at a larger angle, increasing the stress applied to the stellar surface. The stress is minimized when the only inertia coupled to the open field lines is the effective inertia of the electromagnetic field itself. In this sense the estimates which follow are conservative, as mass loading on the field lines will generally increase the magnitude of the spin-down torque or jet power (and reduce the jet's maximum Lorentz factor). On the other hand, the system may behave very differently if the accretion rate is so large that the disk reaches the star and its entire surface is covered with dense plasma; our estimates are not expected to be reliable in this case.

The amount of stellar magnetic flux that is opened following coupling to the disk is difficult to calculate reliably, given uncertainties in the general behavior of MRI turbulence and the specific complications introduced by the stellar field--disk interaction. In the interest of simplicity we therefore subsume the physics that sets the amount of open flux into a model parameter, $\zeta$, and investigate the consequences of its taking different values. The precise value of $\zeta$ can be expected to depend on the conditions in both the accretion disk and the low-density, highly magnetized corona.

If the disk is a good conductor, we can make the estimate that a fraction $\zeta \leq 1$ of the stellar field lines that intersect the disk are opened. If the field is initially dipolar, the poloidal flux function is
\beq
\ff (r,\theta) = \int_0^\theta \int_0^{2\pi} B_r \,r^2\!\sin\!\theta \,{\rm d}\theta\, {\rm d}\phi = 2\pi\mu \frac{\sin\theta}{r}
\eeq
and at the equator $\ff \propto 1/r$. We therefore estimate that when $\rmag < \rlc$ the open flux is given by
\beq
\ffopen = \zeta \frac{\rlc}{\rmag} \ffopeno ,
\label{eq:ffopen}
\eeq
where $\ffopeno$ is the open flux of an isolated star at the same spin frequency (see equation~\ref{eq:n0}); the conducting disk can only increase the open flux and so $\zeta\rlc/\rmag \geq 1$. 
The $\zeta$ parameter describes how efficiently the stellar magnetic field is opened by the star-disk differential rotation, from $\zeta = \rmag/\rlc$ (no disk-induced opening) to $\zeta = 1$ (maximum efficiency or most optimistic case; all stellar field lines which intersect the disk are opened).
The corresponding spin-down torque is
\beq
\Ndownopen = \zeta^2 \lp\frac{\rlc}{\rmag}\rp^2 \No.
\label{eq:nopen}
\eeq
Finding the torque in this way, by replacing $\ffopeno \rightarrow \ffopen$ in equation~(\ref{eq:n0}), should be a good approximation for the fully relaxed state of Fig.~\ref{fig:cartoon}(b);  the poloidal compression of the open flux seen in Fig.~\ref{fig:cartoon}(a) results in a slightly larger torque.

If all of the stellar field is either closed inside $\rmag$ or opened by the disk, the enhanced pulsar wind is the only source of spin-down torque, and so $\Ndown = \Ndownopen$. Likewise, if there are no field lines coupling the star to the disk inside corotation, the spin-up torque outside the accretion column vanishes, $\Nup = 0$. If the amount of flux coupling the star and disk is small (i.e., $\zeta \sim 1$) we can expect $\Nup \ll \Nacc$ and $\Ndown \approx \Ndownopen$; in this approximation we can combine equations~(\ref{eq:nacc}), (\ref{eq:n0}), and (\ref{eq:nopen}) to derive a simple minimal model for the torque, 
\beq
\Ntot = 
  \begin{dcases}
  \mdot\sqrt{GM\rmag} - \zeta^2\frac{\mu^2}{\rmag^2}\frac{\Omega}{c}, & \rmag < \rco \\
  - \zeta^2\frac{\mu^2}{\rmag^2}\frac{\Omega}{c}, & \rco < \rmag < \rlc\\
  - \mu^2 \frac{\Omega^3}{c^3}, & \rmag > \rlc.
  \end{dcases}
\label{eq:ntot}
\eeq
This model is only appropriate for rapidly rotating stars, with spin frequencies greater than roughly a hundred Hz.  For slower pulsars, such as those with periods of around a second, the pulsar wind is much weaker and the spin-down torque will be dominated by those magnetic field lines linking the star to the disk, even if most of the flux is open.

The pulsar wind may manifest itself as a radio jet if it is collimated by external magnetic fields or heavy plasma, such as a disk wind \citep*[e.g.][]{Sulkanen:1990aa,Meier:2001aa}. As field opening adds more flux to the jet, the jet power $\Lj$ will be enhanced by the same factor as the spin-down torque,
\beq
\Lj = \zeta^2 \lp\frac{\rlc}{\rmag}\rp^2 \Lo.
\label{eq:lj}
\eeq

Finally, by changing the relationship between spin frequency and spin-down power the disk-induced opening of magnetic flux will necessitate a modified estimate of the pulsar's magnetic moment. The standard force-free pulsar luminosity for general obliquity is $\Lo = (1+\sin^2\!\theta)\mu^2\Omega^4/c^3$ \citep{Spitkovsky:2006aa}, where $\theta$ is the inclination angle between the magnetic and rotational axes, allowing the magnetic moment to be inferred as $\mu_{\rm std}^2 =  I \nudot c^3/[4\pi^2(1+\sin^2\!\theta)\nu^3]$. If the disk is inside the light cylinder and accretion doesn't occur ($\rco < \rmag < \rlc$), a corrected form should be used; equation~(\ref{eq:lj}) implies 
\beq
\begin{split}
\mu_{\rm corr}& = \zeta^{-1} \frac{\rmag}{\rlc} \, \mu_{\rm std} \\
                      & =  \lp \frac{\xi}{\zeta}\rp^{7/3} \lp 2GM\mdot^2\rp^{-1/3} \ls \frac{I \nudot c}{(1+\sin^2\!\theta)\nu}\rs^{7/6}.
\end{split}
\label{eq:mucorr}
\eeq
Failing to take account of any additional field lines opened by interaction with the disk leads to an overestimate of the stellar magnetic moment.

\begin{figure*}
\includegraphics[width=7.1in, trim = 1mm 2.5mm 2mm 2mm, clip]{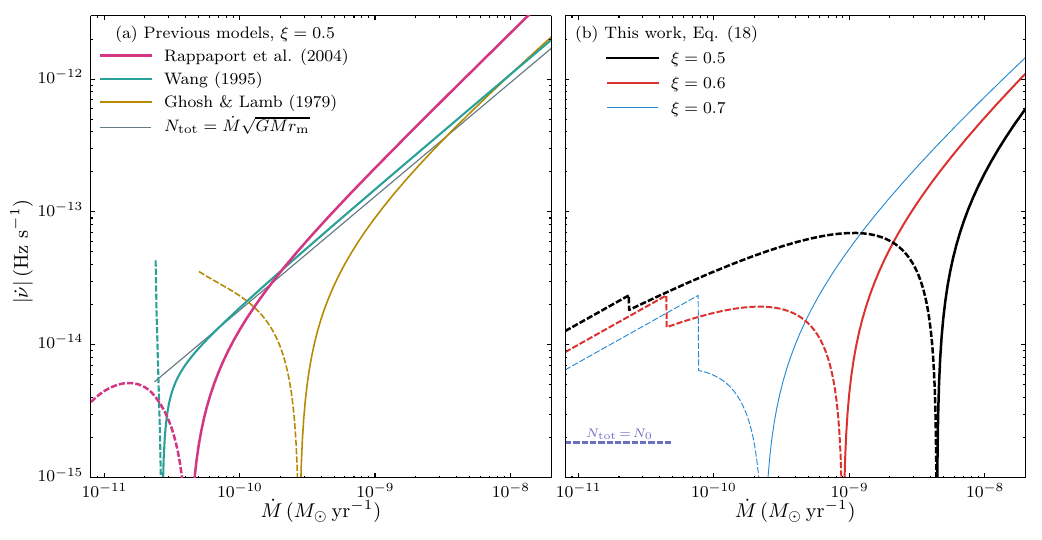}
\caption{\label{fig:nudot} Frequency derivative predicted by various models of the total torque $\Ntot$, for a star with a $10^{26}\Gcmcube$ dipole moment rotating at 500 Hz, as a function of mass accretion rate. (a) Earlier models, all with $\xi=0.5$, including $\Ntot = \Nacc$ for $\rmag < \rco$. (b) Enhanced open-flux model, our equation~(\ref{eq:ntot}) with $\zeta=1$; the spin-down rate due to the standard isolated-pulsar torque $N_0$ is shown in the lower left. Dashed lines indicate a net spin-down torque. The accretion rates at which $\rmag = \rco$, and the accretion torque $\Nacc$ is switched on, correspond to the sharp edges in (b). }
\end{figure*}

\subsection{Numerical estimates}

Using the preceding equations we can estimate the spin-down torque for parameters characteristic of accreting millisecond pulsars in outburst,
\begin{multline}
\Ndown = -8.43 \times 10^{31} \lp\frac{\zeta}{\xi}\rp^2  \lp\frac{\nu}{500\Hz}\rp  \lp\frac{\mu}{10^{26}\Gcmcube}\rp^{6/7}\\
          \times \lp\frac{M}{1.4\msun}\rp^{2/7} \lp\frac{\mdot}{10^{-10}\msunyr}\rp^{4/7} \gcmsq.
\label{eq:ndown_vals}
\end{multline}
For the indicated parameters, $\zeta = 1$, and $\xi = 0.5$,  this gives $\Ndown/\No = 29.3$ (reducing $\nu$ to 250 Hz gives $\Ndown/\No = 117$). 

The jet power can be estimated, using $\Lj = -2\pi\nu\Ndown$, as
\begin{multline}
\Lj = 1.59 \times 10^{36} \lp\frac{\zeta}{\xi}\rp^2  \lp\frac{\nu}{500\Hz}\rp^2  \lp\frac{\mu}{10^{26}\Gcmcube}\rp^{6/7}\\
           \times \lp\frac{M}{1.4\msun}\rp^{6/7} \lp\frac{\mdot}{\mdedd}\rp^{4/7} \ergsec,
\label{eq:lj_vals}
\end{multline}
which appears to be sufficient to power the most powerful neutron star jets (see Section~\ref{obs}); the Eddington accretion rate per solar mass of the central star is $\mdedd = 2.3 \times 10^{-9} \msunyr$, assuming unity radiative efficiency. We also find $\Lj \propto \rmag^{-2} \propto \mdot^{4/7}$; we will discuss this in the light of observations of the radio--X-ray correlation in jet-launching neutron star systems in Section~\ref{obs_jets}.

\subsection{Comparison to other torque models}

In Fig.~\ref{fig:nudot} we show the frequency derivative $\nudot$ of a star, rotating at 500 Hz with a $10^{26}\Gcmcube$ dipole moment, as a function of accretion rate, for five cases in which the total stellar torque is given respectively by $\Nacc = \mdot\sqrt{GM\rmag}$ (equation~\ref{eq:nacc}), the three earlier models which include the disk-magnetosphere interaction (equations~\ref{eq:gl79}, \ref{eq:wang95}, and \ref{eq:rapp04}), and our enhanced pulsar-wind model (equation~\ref{eq:ntot}). Here and in Section~\ref{obs} we relate the torque to a frequency derivative using $\nudot = N/2\pi I$ and a moment of inertia $I = 10^{45} \, \rm{g\, cm^2}$.

Discontinuities are present in the $\nudot$ curves predicted by our model as we have not attempted to smooth the onset of accretion as the magnetospheric radius moves inside corotation---in reality these transitions would be smeared over some range in accretion rate. The edges in Fig.~\ref{fig:nudot}(b) serve to indicate the accretion rates at which $\rmag = \rco$ for the three values of $\xi$ shown; for $\mdot$ values to the left of these edges the system can be said to be in the propeller regime, even though the torque here is due to spin-down on open field lines rather than the ejection of matter from the disk.

At high accretion rates, the spin-up torques predicted by previous models---equations~(\ref{eq:gl79}), (\ref{eq:wang95}), and (\ref{eq:rapp04})---are larger than $\Nacc$ due to the inclusion of positive torque from stellar field lines entering the disk inside $\rco$; in contrast, the $\nudot$ predicted by our model is always less than the standard estimates, and sometimes significantly so. For low accretion rates, where $\rmag < \rco$, our model gives stronger spin-down than equation~(\ref{eq:rapp04}). In our model, not only can the net torque be negative when the star is accreting, but the spin-down can become stronger as $\mdot$ increases, as is demonstrated by the $\xi=0.5$ and 0.6 curves in Fig.~\ref{fig:nudot}(b). The spin-down power on open field lines can increase the critical accretion rate at which the torque goes to zero, especially for smaller values of $\xi$; this will be the subject of the next section. For rapidly rotating stars, the open-flux spin-down torque can potentially have a large effect on the star's spin behavior at all but the highest accretion rates. Finally we note that the new theory is more sensitive than previous models to variations of $\xi$.

\section{Spin equilibrium}
\label{eqlm}

\subsection{Balancing accretion and open-flux torques}

Spin equilibrium is obtained when the total torque on the star vanishes. In the simple model described above (equation~\ref{eq:ntot}) the only spin-up torque is due to the accreting matter and the only spin-down contribution comes from the open field lines, and so equilibrium is found when $\Nacc = -\Ndownopen$,
\beq
\mdot\sqrt{GM\rmag} = -\zeta^2 \lp \frac{\rlc}{\rmag}\rp^2 \No.
\label{eq:spineq}
\eeq
 Using equations~(\ref{eq:nacc}) and (\ref{eq:ndown_vals}) we estimate the equilibrium spin frequency $\nueq$ to be
\begin{multline}
\nueq = 956 \,\zeta^{-2} \xi^{5/2}  \lp\frac{\mu}{10^{26}\Gcmcube}\rp^{-4/7}\\
           \times \lp\frac{M}{1.4\msun}\rp^{1/7} \lp\frac{\mdot}{10^{-10}\msunyr}\rp^{2/7} \Hz,
\label{eq:nueq}
\end{multline}
which is only weakly dependent on the mass accretion rate.

\newpage

\subsection{Maximum steady-state spin frequency for accretion-powered pulsars}

\label{numax}

Equation~(\ref{eq:spineq}) can also be rearranged using equations~(\ref{eq:rlc}--\ref{eq:rm}) and (\ref{eq:n0}) to provide a condition on $\rmag$ (or, equivalently, on $\xi$) which must be satisfied when spin equilibrium is maintained in this way,
\beq
\frac{\rmag}{\rlc} = 2^{-1/2} \frac{\xi^{7/2}}{\zeta^2}.
\label{eq:condition}
\eeq

For X-ray pulses to be detectable during outburst the accretion flow must be held above the stellar surface by the magnetosphere, $\rmag > \rstar$, implying a maximum self-consistent equilibrium spin frequency
\beq
\numax = 3374 \,\zeta^{-2} \xi^{7/2} \lp\frac{\rstar}{10\km}\rp^{-1} \Hz.
\label{eq:numax}
\eeq
Equilibrium via the above mechanism may still be possible if the disk reaches the stellar surface, in which case this constraint is inapplicable. It is likely however that if the accretion rate is significantly larger than the minimal value at which this occurs the open field lines will be flooded with heavy plasma at their bases, and the torque will differ substantially from our estimates.

\begin{figure}
\includegraphics[width=3.4in, trim = 2mm 2mm 2mm 2mm, clip]{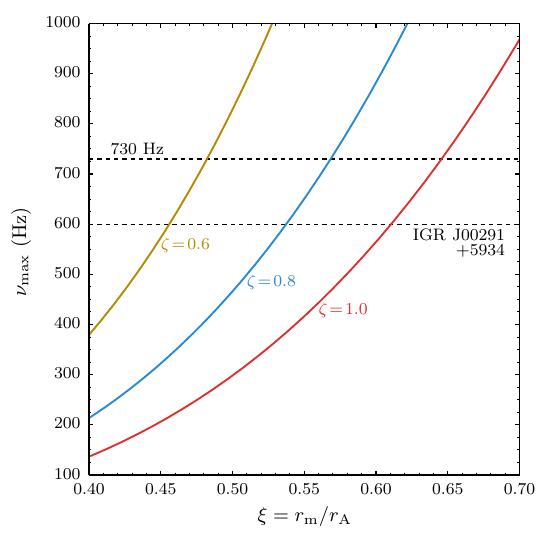}
\caption{\label{fig:numax} Maximum equilibrium spin frequency versus $\xi$ parameter, for three values of $\zeta$ (see equation~\ref{eq:numax}). Also shown is the estimated 730 Hz cut-off to the combined accretion-powered + nuclear-powered pulsar distribution, and the spin frequency of IGR J00291+5934, the fastest accretion-powered pulsar (dashed lines). The model can match the observational cut-off for reasonable $\xi$ and $\zeta$, and would set a maximum steady-state frequency independent of $\mu$ and $\mdot$.}
\end{figure}

Equation~(\ref{eq:numax}) implies that accretion-powered pulsars in instantaneous spin equilibrium are limited to a maximum attainable spin frequency, which is independent of both the stellar magnetic moment and the accretion rate. If $\zeta$ and $\xi$ only vary within narrow ranges, this would enforce a universal cut-off to the spin distribution of accretion-powered pulsars, if these objects can be approximated as being in a steady state at their greatest accretion rate. 

Consider a single star with a particular magnetic moment. There is an accretion rate which places $\rmag$ just above the stellar surface---this is the maximum accretion rate consistent with the star being observable as an accretion-powered pulsar, i.e. with pulses formed by magnetically confined accretion columns. This accretion rate also sets the star's maximum steady-state spin frequency via equation~(\ref{eq:spineq}); if the spin frequency were larger, the star would spin-down towards $\numax$ (see Section~\ref{attractor}). If $\mdot$ were to increase, the star's steady-state spin frequency would increase, but it would no longer be observed as an accretion-powered pulsar. Equation~(\ref{eq:numax}) states that the maximum spin frequency found in this way is universal for stars of all magnetic moments.

In Fig.~\ref{fig:numax}, $\numax$ is shown as a function of the model parameters. The maximum frequency can be made to coincide with the observationally estimated cut-off of $\sim 730$ Hz for reasonable values of $\xi$, provided that $\zeta$ is large enough (i.e., that field lines are opened efficiently by the disk).

\subsection{Other conditions satisfied by equilibrium solutions}

In this section we describe several additional relationships which must hold for a spin equilibrium solution to be physically valid, and show that these conditions are satisfied by the observed systems given reasonable ranges of the model parameters.

The star must be accreting for spin equiilibrium to be possible, and so we can demand that the magnetospheric radius is inside the corotation radius, $\rmag < \rco$; equations~(\ref{eq:rco}) and (\ref{eq:condition}) then imply that in equilibrium the star must be spinning faster than some minimum frequency, $\nueq > \numin$, where
\beq
\numin = 8160 \,\zeta^{-6} \xi^{21/2} \lp\frac{M}{1.4\msun}\rp^{-1} \Hz.
\label{eq:numin}
\eeq

The condition that $\rmag < \rco$ in spin equilibrium can also be understood as requiring that $\rmag$ lies inside $\rmagmax$, the largest possible corotation radius consistent with equilibrium at a given $\zeta$ and $\xi$, which is 
\beq
\rmagmax = \rco(\numin)  = \zeta^4 \xi^{-7} r_{\rm s}, 
\label{eq:rmagmax} 
\eeq
where $r_{\rm s} = 2 G M/c^2$ is the star's Schwarzschild radius. Demanding $\rmag < \rmagmax$ in turn implies that $\mdot > \mdot_{\rm min}$,
\begin{multline}
\mdot_{\rm min} = 1.81\times10^{-7}  \zeta^{-14} \xi^{28}  \\ \times \lp\frac{\mu}{10^{26}\Gcmcube}\rp^2 \lp\frac{M}{1.4\msun}\rp^{-4}  \msunyr.
\end{multline}
Although the very strong dependence on the $\zeta$ and $\xi$ parameters makes reliance on this quantity dubious, we can see that for, say, $\zeta = 0.9$ and $\xi = 0.7$ one finds $\mdot_{\rm min} = 3.6\times10^{-11} \msunyr$, distinctly lower than the observed outburst accretion rates for the millisecond pulsars; as $\xi$ is reduced further, $\mdot_{\rm min}$ becomes small very rapidly.

Clearly, we require $\numin < \numax$; since $\numin/\numax \approx 2.42\, \zeta^{-4}\xi^{7}$ for our standard stellar mass and radius, this provides a maximum self-consistent value of $\xi$ for a given $\zeta$ in spin equilibrium,
\beq
\xi < 0.881 \, \zeta^{4/7}.
\label{eq:ximax}
\eeq

If we wish to explain the observed accretion-powered millisecond-pulsar spin distribution as consisting of systems in approximate equilibrium, we can demand $\numax > 730\Hz$, implying a minimum $\xi$,
\beq
\xi > 0.645 \lp \frac{\numax}{730 \Hz}\rp^{2/7} \zeta^{4/7}.
\label{eq:ximin}
\eeq
Note that for $\numax = 730 \Hz$ the ratio of minimum to maximum allowable $\xi$ is comfortably smaller than unity, independent of $\zeta$. 

Smaller values of $\xi$ are required to explain slower equilibrium frequencies in this framework, because for slower spin one must open more field lines for a given accretion rate (i.e.\ $\Ndown \propto \xi^{-2} \nu$; see equation~\ref{eq:ndown_vals}). It is not clear at what frequency the enhanced pulsar wind would cease to be the dominant source of spin-down torque, but using equation~(\ref{eq:numin}) we can estimate an alternative maximum $\xi$,
\beq
\xi < 0.658 \lp \frac{\numin}{100 \Hz}\rp^{2/21} \zeta^{12/21}.
\label{eq:ximax2}
\eeq
The very weak sensitivity to $\numin$ implies that $\xi \sim 0.5$ would provide a valid equilibrium solution even for slow rotators. 

Equations~(\ref{eq:ximax}--\ref{eq:ximax2}) imply that $\xi$ varying within the reasonable range $0.5 \lesssim \xi \lesssim 0.8$ would produce valid equilibrium solutions for the entire millisecond pulsar distribution. Values of $\xi$ covering much of this range (in particular $0.55 \leq \xi \leq 0.72$) were measured in a numerical study of the dependence of $\rmag$ on $\mu$ and $\mdot$ by \cite{Kulkarni:2013aa}.

\subsection{Relating equilibrium rotation rate to magnetospheric radius}
\label{attractor}

\begin{figure}
\includegraphics[width=3.4in, trim = 2mm 2mm 2mm 2mm, clip]{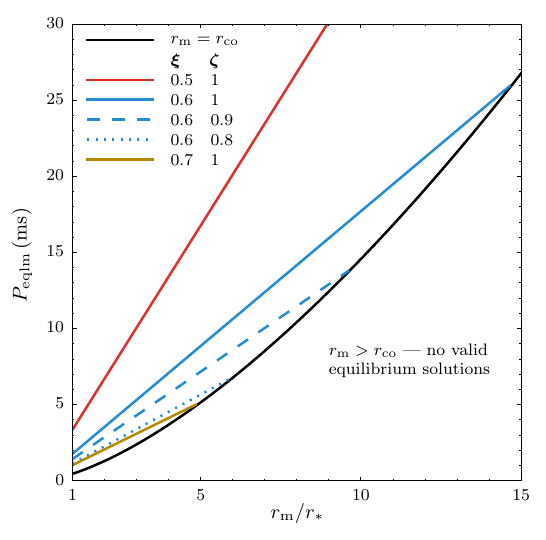}
\caption{\label{fig:eqcurves} Equilibrium spin period versus magnetospheric radius for the standard model (black curve) and our enhanced pulsar wind model (colored lines). The equilibrium described here can only hold for $\rmag < \rmagmax$, where $\rmagmax$ is given by the intersection of a particular $\peq(\zeta,\xi)$ line with the $\rmag=\rco$ curve.}
\end{figure}

We could alternatively think of equation~(\ref{eq:condition}) as expressing the equilibrium period $\peq$ predicted by the model,
\beq
\peq = \frac{2^{3/2}\pi}{c} \zeta^2 \xi^{-7/2} \rmag ;
\label{eq:peq}
\eeq
here $\peq \propto \rmag$. In the standard picture of spin equilibrium, $\rmag \approx \rco$ \citep[e.g.][]{Wang:1995aa} and so $\peqstd \approx 2\pi (GM)^{-1/2} \rmag^{3/2}$. These two scenarios are illustrated in Fig.~\ref{fig:eqcurves}. In this diagram a star spinning up at constant accretion rate moves vertically downward, assuming $\xi$ remains constant. 

If $\rmag < \rmagmax$ (see equation~\ref{eq:rmagmax}), the star will eventually reach spin equilibrium at the period given by equation~(\ref{eq:peq}). The torque at any stage is determined by the competition between $\Nacc$, which is not a function of the spin period, and $\Ndown \propto P^{-1}$. When $P > \peq$, $\Nacc + \Ndown > 0$ and the star spins up, decreasing $P$ and increasing the strength of the spin-down torque until an equilibrium is reached. Likewise when $P < \peq$, $\Nacc + \Ndown < 0$ and the star's period increases, decreasing the spin-down strength until $\Nacc+\Ndown=0$. In this way equation~(\ref{eq:peq}) describes an attractor for $P$.

If $\rmag > \rmagmax$, the corotation radius, which moves inward as the star is spun up, reaches $\rmag$ before the star is rotating fast enough to establish spin equilibrium using the open flux set by $\rmag$. In this case the star's rotation will hover around $\peqstd$: accretion is shut off when $\rco < \rmag$, the star spins down, $\rco$ increases, until accretion and spin-up restart \citep[see e.g.][]{Spruit:1993aa,DAngelo:2010aa}. This long-term averaged (as opposed to instantaneous) spin equilibrium may operate in systems hosting neutron stars with longer periods. 

Fig.~\ref{fig:eqcurves} suggests that accretion torques may be instantaneously balanced by spin-down torques on open field lines for some of the most rapidly rotating stars, in the most extreme cases preventing spin-up to sub-millisecond periods.

\begin{figure}
\includegraphics[width=3.4in, trim = 2mm 2mm 2mm 2mm, clip]{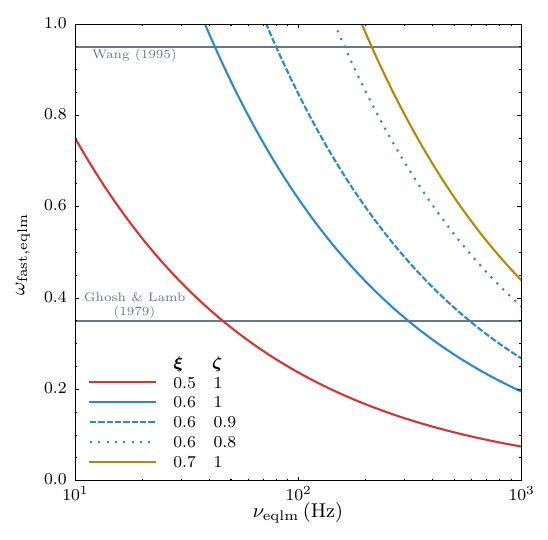}
\caption{\label{fig:fastness} Fastness parameter $\wfast = \Omega/\Omega_{\rm K}(\rmag)$ versus spin frequency, at spin equilibrium. For a valid equilibrium solution $\wfasteq$ must be less than unity.}
\end{figure}

The \cite{Ghosh:1979aa} and \cite{Wang:1995aa} torque models, equations~(\ref{eq:gl79}) and (\ref{eq:wang95}), predict fastness parameters at spin equilibrium (often referred to as the critical fastness parameter) of $\wfasteq = 0.35$ and 0.95 respectively, independent of $\nueq$. Using equations~(\ref{eq:rco}) and (\ref{eq:condition}) we can express this quantity in our model as 
\beq
\wfasteq = 2^{-3/4} \zeta^{-3} \xi^{21/4} c^{3/2} \lp G M \Omega \rp^{-1/2}.
\label{eq:wfasteq}
\eeq
In Fig.~\ref{fig:fastness}, $\wfasteq$ is plotted against $\nueq$ for several combinations of $\zeta$ and $\xi$. A valid spin equilibrium solution requires $\rmag < \rco$ and hence $\wfasteq < 1$; the locations where $\wfasteq = 1$ therefore coincide with the minimum allowable spin frequencies of equation~(\ref{eq:numin}). As the equilibrium frequency increases, the larger spin-down power must be balanced by greater mass accretion and hence a decreasing magnetospheric radius, which is reflected in $\wfasteq$ being a decreasing function of $\nueq$.

\section{Comparison to Observations}
\label{obs}

In this section we express magnetic moments as $\mu = \mu_{26}  \times 10^{26}\Gcmcube$, accretion rates as $\mdot = \mdot_{-10}\times10^{-10}\msunyr$, torques as $N = N_{32} \times 10^{32} \gcmsq$, and frequency derivatives as $\nudot = \nudot_{-14} \times 10^{-14} \Hzs$. Where necessary we convert reported surface magnetic field estimates $B$ to dipole moments using $\mu \sim B \rstar^3$; when spin-down frequency derivatives in quiescence are available we calculate $\mu$ directly using the force-free aligned rotator formula.

\subsection{Spin frequency distribution of millisecond \\ X-ray pulsars}

The frequency distribution of rapidly rotating pulsars has been puzzling at least since \cite{White:1997aa} reported an apparent correlation between magnetic field strength and accretion rate, required to explain the existence of pulsars of similar, presumably equilibrium, periods at widely varying accretion rates. This question led \cite{Bildsten:1998aa} to suggest that the strongly frequency-dependent onset of spin-down via gravitational radiation enforces a maximum cut-off rotation rate; alternatively, the spin-up torques may be reduced at high accretion rates due to the formation of a radiation-pressure supported sub-Keplerian inner disk \citep{Andersson:2005aa}. While the discovery of more objects has obviated concerns over a magnetic field--X-ray luminosity correlation \citep{Patruno:2012ab,Papitto:2014ac}, the problem of explaining the lack of pulsars above $\sim 700$ Hz remains unresolved.

We described in Section~\ref{theory} how interaction with a disk leads to increased open stellar magnetic flux, allowing a state where the spin-up accretion torque is instantaneously balanced by angular momentum loss through the pulsar wind, as outlined in Section~\ref{eqlm}. For magnetic moments and accretion rates consistent with observations, $\mu_{26} \sim 1$ and $\mdot_{-10} \sim 1$--50, and theoretically reasonable model parameters $\zeta \lesssim 1$ and $\xi \sim 0.5$--0.8, the model can explain the spin frequencies of the observed millisecond X-ray pulsars as being set by the termination of spin-up by the star entering the spin equilibrium state given by equations~(\ref{eq:spineq}) and (\ref{eq:nueq}). 

In particular, the model predicts a maximum steady-state spin frequency for sources which appear as accretion-powered pulsars, as set by equation~(\ref{eq:numax}); see Fig.~\ref{fig:numax}. This cut-off is independent of the accretion rate and the star's magnetic field, and would therefore be universal if all magnetosphere-disk interactions can be described by similar $\xi$ and $\zeta$; it can be reconciled with the observationally estimated $\sim 730$ Hz maximum frequency for the same reasonable range of model parameters as above.

A model in which the primary limitation on an accreting star's rotation rate is due to magnetically applied torques would accord with observations which suggest that the nuclear-powered pulsars (burst oscillation sources) have a faster spin distribution than both the accretion-powered and radio millisecond pulsars \citep{Papitto:2014ac}. The nuclear-powered pulsars generally have a weaker surface magnetic field than the accretion-powered pulsars, giving a smaller spin-down torque for the same accretion rate at all stages of an outburst, and possibly periods of little or no spin-down torque when the stellar surface is flooded at the highest accretion rates. These objects are not strictly subject to the high-frequency cut-off of Section~\ref{numax}, but their maximum attainable spin frequency can be expected to be similar to that of the accretion-powered sources, as appears to be the case, if the magnetic field strengths of the two populations don't differ too substantially. The frequencies of the nascent radio millisecond pulsars may have been systematically reduced by the spin-down mechanism described here during the terminal, low accretion rate, phases of Roche-lobe overflow, in a manner analogous to \cite{Tauris:2012aa}, who used a variant of equation~(\ref{eq:rapp04}) for the propeller-regime spin-down torque.

Another possible scenario is that the star's increasing spin frequency may not be able to reach the equilibrium frequency commensurate with the outburst's maximum accretion rate before the end of an outburst, and the frequency increase sustained by the star may be approximately balanced by strong net spin-down due to the enhanced pulsar wind as the accretion rate declines; see Fig.~\ref{fig:nudot}. In this way the stellar spin frequency may oscillate in a sawtooth pattern over the outburst-recurrence timescale while being in a kind of long-term spin equilibrium.

\subsection{Spin evolution of individual sources}

For several sources, frequency measurements during multiple accretion episodes allow the spin-down rate during X-ray quiescence to be found, with which one can estimate the star's magnetic moment using the force-free spin-down formula. This in turn permits tests of the torque models during, or integrated over, a single accretion outburst. This is complicated by the presence of strong timing noise in the observations, making the calculation of reliable $\nudot$'s much more problematic than for isolated pulsars, and the nonlinear response of both the accretion and spin-down torques to the accretion rate. Despite these difficulties, a concern has been voiced that some of these objects do not show as strong a spin-up during outbursts as the X-ray luminosity would imply under the standard models \citep[e.g.][]{Patruno:2012aa}; in some cases no change in frequency was discernable despite the inference of a large accretion rate. Here we briefly show that the anomalous spin-down torque required is approximately consistent with the torque model of equation~(\ref{eq:ntot}) for reasonable choices of the various parameters, and describe similar observations of other systems which do not require additional spin-down. We take $\Ntot = \Nacc = \mdot\sqrt{GM\rmag}$ as the standard torque model at high accretion rates (see Fig.~\ref{fig:nudot}).

An RXTE study of the pulse frequency during five outbursts of the $\nu = 401$ Hz pulsar SAX J1808.4-3658 by \cite{Hartman:2008aa, Hartman:2009aa} has revealed a long-term spin-down rate of $\nudot_{-14} \approx -0.055$. This spin-down is consistent with being constant in time, and is too large to be due to torques during the outbursts given their duration and the single-outburst upper limit of $|\nudot_{-14}| < 2.5$, and so is assigned to spin-down during X-ray quiesence; this implies a dipole moment of $\mu_{26} \approx 0.76$ (without torque enhancement). 

Using an average outburst accretion rate of $\mdot_{-10} \approx 5$ \citep{Haskell:2011aa}, the spin-up torque in the standard theory would be $\Nacctt \approx 5.9  \sqrt{\xi}$, implying a frequency derivative $\nudot_{-14} \approx 9.4\sqrt{\xi}$, well above the observational upper limit. The spin-down torque, equation~(\ref{eq:nopen}), would be $\Ndowntt \approx 1.4(\zeta/\xi)^2$ and so the combined model, equation~(\ref{eq:ntot}), predicts 
$\nudot_{-14} \approx 9.4 \sqrt{\xi} - 2.2 (\zeta/\xi)^2$.
For the frequency derivative in the model to obey the observational constraint (i.e., $\nudot_{-14} < 2.5$), one must therefore have $\xi <$ [0.65, 0.61, 0.55] for models in which $\zeta =$ [1.0, 0.9, 0.8] respectively; note that $\rmag > \rstar$ for these values of $\xi$. As these values of $\xi$ are reasonable, given previous theoretical and numerical studies, the enhanced open-flux model can explain the undetectable spin-up of SAX J1808.4-3658 if $\zeta$ is sufficiently close to unity.

The long-term mass transfer rate in the J1808 system is expected to be approximately $10^{-11} \msunyr$ \citep{Bildsten:2001aa, Heinke:2009aa}; for this value the corrected magnetic moment given by equation~(\ref{eq:mucorr}) would be $\mu_{\rm corr, 26} \approx 0.15$, which is inconsistent with the detection of pulsations at high accretion rates \citep{Psaltis:1999aa, Hartman:2008aa}. This may be resolved by a more detailed model of the build up of the disk between outbursts.

Observations of the 2003 outburst of the 314 Hz pulsar XTE J1814-338 placed an upper limit of $|\nudot_{-14}| < 1.5$ on the frequency derivative \citep{Haskell:2011aa}. As only one outburst has been detected by an instrument with sensitive timing capability, no estimate can be made of the quiescent spin-down rate and hence magnetic moment. Using the average outburst accretion rate of $\mdot_{-10} \approx 2$ \citep{Haskell:2011aa} one derives, in the standard torque model, an expected spin-up rate of $\nudot_{-14} \approx 4.7 \mu_{26}^{2/7} \sqrt{\xi}$, well above the observational constraint for $\mu_{26} \sim 1$. Including spin-down torques on open field lines one finds 
$\nudot_{-14} \approx 4.7  \,\mu_{26}^{2/7} \sqrt{\xi} - 1.3 \,\mu_{26}^{6/7} (\zeta/\xi)^{2}$,
giving, for $\mu_{26} = 1$, scenarios satisfying the observational constraints (i.e., $|\nudot_{-14}| < 1.5$) when $\xi < [0.72, 0.67, 0.61, 0.56]$ for $\zeta = [1.0, 0.9, 0.8, 0.7]$ respectively; all of these $\xi$ result in $\rmag>\rstar$ as $\ra = 28.92\, \mu_{26}^{4/7}\km$ for this accretion rate. Note that this source can be in spin equilibrium during such an outburst given spin-down on open flux alone: $\nudot = 0$ for $\xi = [0.59, 0.54, 0.49]$ using $\zeta = [1.0,0.9,0.8]$. We conclude that an enhanced open-field torque may be responsible for the weak or absent spin-up of XTE J1814-338.

Two systems for which accretion-induced spin-up has been measured do not appear to require enhanced braking torque. The 2002 outburst of the 435 Hz pulsar XTE J1751-305 had a peak X-ray luminosity of $L_{\rm X} \approx 2.7\times 10^{37} \ergsec$ for an estimated distance of 8.5 kpc \citep{Markwardt:2002aa, Gierlinski:2005aa}, implying a mass accretion rate of $\mdot \approx 2.3 \times 10^{-9} \msunyr \approx \mdedd$. The peak frequency derivative has been reported as $\nudot_{-14} \approx 56$ by \cite{Papitto:2008aa}. The long-term spin-down rate of $\nudot_{-14} \approx -0.55$ indicates a magnetic moment of $\mu_{26} \approx 2.1$ \citep{Riggio:2011aa}. The above values give an outburst peak frequency derivative, due to the spin-up torque component $\Nacc$ alone, of $\nudot_{-14} \approx 47 \sqrt{\xi}$, smaller than the measured value, and therefore additional spin-down torque is not called for.

A large spin-up during outburst has also been observed for IGR J00291+5934 \citep{Falanga:2005aa}, at 599 Hz the fastest rotating accretion-powered pulsar. The quiescent spin-down of $\nudot_{-14} \approx - 0.41$ \citep{Hartman:2011aa, Papitto:2011aa} gives $\mu_{26} \approx 1.1$. \cite{Patruno:2010aa} estimated the peak accretion rate of the 2004 outburst to be $\mdot_{-10} \sim 20$ and measured a spin-up rate of $\nudot_{-14} \approx 51$. The accretion spin-up torque alone would give $\nudot_{-14} \approx 35 \sqrt{\xi}$ for the above parameters, or $\nudot_{-14} \approx 42$ if the torque were applied from the corotation radius. No additional spin-down torque component is required to explain the timing behavior of this system in outburst.

\subsection{Jets from neutron star X-ray binaries}
\label{obs_jets}

The radio emission of Cir X-1 and Sco X-1 has been resolved into clearly defined jets. The brightening of the Cir X-1 radio lobes following an X-ray flare implies energy transfer at Lorentz factor $\Gamma > 15$, and minimal-energy arguments for powering the extended emission require a jet power $\Lj \gtrsim 10^{35} \ergsec$ \citep{Fender:2004aa}. For Sco X-1, energy transfer along the jet occurs at $\Gamma > 3$, and equipartition magnetic fields of $\sim 0.3$--$1 \G$ and a radiative lifetime of $\sim 30$ minutes again demand a jet power in excess of a few $10^{35} \ergsec$ \citep{Fomalont:2001ab, Fomalont:2001aa}. These objects typically accrete at a large fraction of the Eddington rate, and are categorized as Z sources in the spectral and timing classification of \cite{Hasinger:1989aa}. Both may be rapidly rotating, as suggested by separations of a few hundred Hz between their twin kHz quasi-periodic oscillations \citep{van-der-Klis:1996aa, Boutloukos:2006aa}, although see \cite{Abramowicz:2003aa} for an alternative viewpoint on the kHz QPOs. It has been proposed that the Z sources have relatively strong magnetic fields; dipole moments in excess of $10^{27} \Gcmcube$ have been estimated for Cyg X-2 \citep{Focke:1996aa} and XTE J1701-462 \citep{Ding:2011aa}. 

If the magnetic moments of Cir X-1 and Sco X-1 are sufficiently large, greater than about $10^{26} \Gcmcube$, the required jet powers can be supplied by the rotating open stellar field lines whose number has been increased by interaction with the surrounding disk. For $\mu_{26} = 1$, $\nu = 300 \Hz$, and mass accretion at half the Eddington rate (i.e. $\mdot = 0.5 [M/\!\msun] \mdedd$) equation~(\ref{eq:lj_vals}) gives $\Lj = 4.6 \times 10^{35} (\zeta/\xi)^2 \ergsec$ (for $\zeta/\xi = 1$ this is two orders of magnitude larger than the spin-down power of an isolated pulsar of the same dipole moment and spin).

When the star's spin and magnetic axes are nearly aligned, the center of the open flux bundle will be magnetically shielded from the accreting matter, allowing the formation of a Poynting flux-dominated jet with low baryon loading and hence high terminal Lorentz factor; a similar effect was observed in the non-relativistic propeller-regime simulations of \cite{Romanova:2009aa}.

Several accreting millisecond pulsars, including SAX J1808.4-3658 and IGR J00291+5934, are associated with radio emission which is attributed to a jet; they have generally been found to have stronger radio emission than the atoll sources at the same X-ray luminosity \citep*{Migliari:2011aa}. This could be explained in the open stellar field framework by their having generally higher spin frequencies or stronger magnetic fields. The latter may be the case, as accretion-powered pulsars need relatively strong fields to confine the accretion column and atolls are thought to be weak-field objects; for example the magnetic moment of the atoll source Aql X-1 has been estimated to be less than $10^{26}\Gcmcube$ \citep{Zhang:1998aa, Campana:2000aa}. Analyses of the relative rankings of sources' rotation rates and radio emission have found some evidence for higher spin frequency being associated with higher radio luminosity, as would be expected in a model where the jet power ultimately derives from stellar rotation \citep{Migliari:2011aa, King:2013aa}.

The most basic model of jets powered by the enhanced stellar open flux, equation~(\ref{eq:lj_vals}), predicts a comparatively weak scaling of jet power with accretion rate, $\Lj \propto \mdot^{4/7}$. This is similar to what is seen in Aql X-1, a rapidly rotating star with 549 Hz type I X-ray burst oscillations \citep{Zhang:1998ab}, whose radio luminosity $\Lr$ scales with X-ray luminosity as $\Lr \propto \Lx^{0.6}$; on the other hand, 4U 1728-34 shows a stronger dependence, $\Lr \propto \Lx^{1.4}$ \citep{Migliari:2006aa}, which is consistent with $\Lj \propto \mdot$ in the scale-invariant disk-jet model \citep{Heinz:2003aa}. In the scenario we describe here, $4/7$ is a lower limit to the scaling power with the accretion rate, since we base the jet power on the spin-down luminosity of the standard isolated pulsar wind solution, which is the lowest spin-down state for a given open magnetic flux. If the disk is thick the poloidal flux will be pushed away from the equator, increasing the jet power; as $\mdot$ increases, the magnetospheric radius moves inward and the amount of poloidal field line displacement increases, which corresponds to the jet power scaling somewhat more strongly with $\mdot$.

Neutron star X-ray binaries have generally shown less tendency than those with black holes to suppress the jet in the soft X-ray state \citep{Migliari:2006aa}. In the model discussed here, jet suppression would be expected when the accretion rate increases to the point that the disk reaches the stellar surface and all of the field lines are inundated with accreting matter. 

Quenching has been observed in Aql X-1 at accretion rates above about $0.1 \,\mdot_{\rm Edd}$ \citep{Tudose:2009aa,Miller-Jones:2010aa}. At this accretion rate the Alfv\'{e}n radius $\ra \approx 10\km$ for a dipole moment of $\mu_{26} = 0.2$, so if the field is relatively weak, $\mu_{26} \lesssim 0.7$, flooding may be responsible for the jet suppression, depending on the effective $\xi = \rmag/\ra$. Neutron stars with stronger fields may be immune to this type of quenching; for $\mu_{26} > 2$ the Alfv\'{e}n radius is above 20 km for accretion at the Eddington rate, and so the jet should persist even at the highest accretion rates unless $\xi$ is unexpectedly small. This mechanism would explain why only some binary systems, those with stars whose magnetic moments are smaller than a critical value of roughly a few $10^{26}\Gcmcube$, are subject to jet suppression in the high accretion rate, soft X-ray state. This contrasts with jet models based solely on accretion disk physics, similar to those employed for accreting black holes, where quenching may be tied to a transition from a thick to a thin disk at roughly $0.1 \,\mdot_{\rm Edd}$; in that case all systems should show jet suppression in the luminous soft state.

Several points in the above discussion would be clarified by reliable measurements of the dipole moments of jet-associated neutron stars, particularly as the proposed difference in field strength between the Z and atoll sources has been challenged by observations of XTE J1701-462, which has been observed to display both Z- and atoll-like behavior during a single outburst \citep{Lin:2009aa}.

\section{Discussion \& Conclusion}
\label{discussion}

The rotational energy of rapidly spinning neutron stars can be efficiently extracted in a powerful Poynting flux-dominated wind, even for stars with comparatively low magnetic dipole moments, if magnetic field lines which would otherwise close within the light cylinder can be forced to open to infinity; the strength of the usual electromagnetic pulsar wind is effectively boosted by a large factor. A natural means of opening these field lines is coupling to a surrounding accretion disk---differential rotation between the footpoints of a field line linking the stellar surface to a Keplerian disk causes toroidal magnetic field to accumulate and the field line to inflate outward, eventually achieving an open-field configuration. 

We have presented a simplified minimal model of the torque on the star and the power in the electromagnetic wind when a fraction of the total stellar flux which would otherwise link to the disk is opened. This wind may be collimated by external gas or magnetic pressure, or by mass loading on its outermost magnetic field lines, creating a jet. We have applied our torque and jet power estimates to three problems: 

\begin{enumerate}[leftmargin=*]
\item \emph{The millisecond X-ray pulsar spin-frequency cut-off.} We find that at high spin frequency, and moderate accretion rates and dipole moments, the spin-up torque due to accretion can be balanced by the spin-down torque from the open field lines alone, giving spin equilibrium and preventing spin-up to sub-millisecond periods. Unlike the proposed action of gravitational radiation, in this case the torque does not scale as a high power of the frequency and so would not enforce a nearly universal maximum spin. Nevertheless, the model predicts a maximum steady-state spin frequency for neutron stars which are observable as accretion-powered pulsars, a limit which is independent of accretion rate and stellar magnetic field strength. For individual sources below the cut-off, the equilibrium spin frequency depends on parameters like the accretion rate and the stellar mass and magnetic field, which appears to be generally consistent with the observed flat spin distribution of accreting millisecond pulsars.

\item \emph{Weak or absent spin-up of millisecond pulsars in outburst.} Accretion episodes of two sources, SAX J1808.4-3658 and XTE J1814-338, resulted in undetectable frequency increases, despite X-ray emission which would lead one, using standard accretion torque theory, to expect spin-ups well above the observational limits. We have argued that the enhanced open flux spin-down torque could have been strong enough to prevent measurable spin-up in these systems. 

\item \emph{Relativistic jets from neutron star binaries.} A model in which the jet is powered by stellar rotation and mediated by enhanced open stellar flux can account for the strengths of the Cir X-1 and Sco X-1 jets, for the observed accretion rates and reasonable magnetic dipole moments and spin frequencies. The field lines at the center of the jet would not connect to the disk and would be shielded from contaminating baryons, which may result in the observed high jet Lorentz factors. This model also appears to be consistent with the higher radio luminosity of the accreting millisecond pulsars compared to the atoll sources, and may explain the jet suppression in Aql X-1 at high accretion rates and the lack of such quenching in other sources.
\end{enumerate}

The problem of accreting magnetic stars is clearly highly complicated, and the idealized picture described here can only possibly tell a small part of the story. Our basic conclusion is that the open stellar magnetic flux could make an important contribution to the behavior of many neutron star systems if its share of the total stellar flux can be increased by the presence of surrounding plasma. We will present results of numerical simulations addressing various aspects of this problem  in the context of our model in a future paper (K. Parfrey et al. 2016, in preparation).

For simplicity we have assumed that the stellar magnetic field and the accretion disk are in a stationary configuration, determined by the accretion rate which can be taken as constant; this is only a rough approximation, but may often be acceptable, as the timescale for relaxation of the magnetosphere to the steady state will be related to the spin period of the star, which is very short for the objects of interest. There are two cases where time dependence may be particularly relevant. Firstly, in the early stages of accretion outbursts, when the inflating stellar magnetic field is competing with the increasing accretion rate and more flux may couple the star and disk. In this case the spin-up torque contribution will be higher and the spin-down contribution lower than in our estimates, leading to stronger spin-up; this may be responsible for the large frequency derivatives observed over outbursts of XTE J1751-305 and IGR J00291+5934. Secondly, increases in the accretion rate or changes in the disk structure coincident with X-ray binary state transitions may trigger the sudden opening of additional magnetic field lines, leading to reconnection and the emission of large plasmoids before the steady state is reached; this may be related to observations of powerful transient jets associated with state transitions \citep{Migliari:2006aa}.

Magnetic flux connecting the star to disk plasma with lower angular velocity applies a stronger spin-down torque to the star than the same flux opened to infinity and contributing to spin-down via the enhanced pulsar wind, if the disk's resistivity is low and field lines do not easily slip through it. That open field lines can generate less spin-down torque than closed ones led \cite{Matt:2005aa} to worry that a largely open configuration would have difficulty preventing accreting protostars from spinning up to unobserved angular frequencies. We also propose that there will generally be little flux coupling the star and disk, however in the case of rapidly rotating neutron stars the combination of high spin and relatively strong magnetic fields means the open flux can nevertheless dominate the system energetically. For these objects external mass loading is unnecessary because the rotationally induced voltages are large enough for magnetospheric pair creation to supply the charge carriers required by the electromagnetic wind, as manifested in the coherent radio emission of isolated pulsars of similar spins and magnetic moments; additional inertia on the open field lines would however increase a star's spin-down power.

The picture of neutron star jets proposed here is related to Blandford-Znajek-type models for black hole jets, in that in both cases the jet power is extracted from the rotational energy of the compact object by helical magnetic field lines; it makes little qualitative difference whether the power transfer from the compact object to the field is mediated by a rotating space-time or a rotating solid surface. This outlook contrasts with magnetocentrifugal jet models, in which the jet is launched by field lines attached to the inner accretion disk; it has been argued that to produce a jet in this way the stellar magnetic moment must be small enough that the disk extends down to the stellar surface \citep{Massi:2008aa}.  Both mechanisms may operate in the same system at different accretion rates or even concurrently---in particular a disk-launched jet may persist when the stellar-field jet is quenched at high accretion rates.

The lack of observed jets from strong-field pulsars in binaries \citep[e.g.][]{Fender:1997aa} may also be explained in this framework---here jet power scales like $\Lj \propto \mu^{6/7} \nu^{2}$, and strong-field pulsars' high dipole moments are outweighed by their slower rotation, to which the jet power is more sensitive; the jet power from a star with $B \sim 10^8 \G$ and $\nu \sim 500 \Hz$ is about 100 times larger than from one with $B\sim 10^{12} \G$ and $\nu \sim 1 \Hz$, for the same accretion rate and $\zeta/\xi$.

An interesting new class of objects is the transitional millisecond pulsars: neutron stars which switch between X-ray pulsar and radio pulsar phases, over timescales as short as weeks \citep[e.g.][]{Archibald:2009aa, Papitto:2013aa, Bogdanov:2014aa}. Of these sources, only for PSR J1023+0038 has a magnetic moment been estimated by measuring a frequency derivative in the radio phase \citep{Archibald:2013aa}, and there are no constraints on frequency derivatives during accretion episodes. The increase of the apparent dipole moment inferred from spin-down measurements due to the opening of additional stellar flux (see equation~\ref{eq:mucorr}) may be particularly important for the transitional pulsars, whose pulsed radio emission may resume before the remnant disk is cleared from the light cylinder. It is hoped that the unique possibilities of these systems may prove useful for disentangling the relationships between accretion rate, stellar magnetic field, spin-up/spin-down torques, and jet launching.

\vspace{0.7cm}
KP is supported by the Max-Planck/Princeton Center for Plasma Physics.  AS acknowledges support from NASA grants NNX14AQ67G, NNX15AM30G, and a Simons Investigator Award from the Simons Foundation.  AMB acknowledges support from NASA grant NNX13AI34G.

\bibliographystyle{apj}
%\bibliography{bib_link.bib}

\end{document}